\documentclass[a4paper,aps,pra,twocolumn,showpacs,preprintnumbers,amsmath,amssymb]{revtex4}

\usepackage{bbm}
\usepackage{amsmath}
\usepackage{verbatim}
\usepackage{graphicx}
\usepackage{epsfig}
\usepackage{times}

\newcommand{\beq}{\begin{equation}}
\newcommand{\eeq}{\end{equation}}
\newcommand{\beqa}{\begin{eqnarray}}
\newcommand{\eeqa}{\end{eqnarray}}

\begin{document}
\title{Can the states of the W-class be suitable for teleportation}

\author{V.N. Gorbachev,\footnote{E-mail: vn@vg3025.spb.edu}
A.I. Trubilko, A.A. Rodichkina}

\affiliation{Laboratory of Quantum Information and Computing,
University of AeroSpace Instrumentation, 67, Bolshaya Morskaya,
St.-Petersburg, 190000, Russia}

\begin{abstract}
Entangled states of the W-class are considered as a quantum
channel for teleportation or the states to be sent. The protocols
have  been found by unitary transformations of the schemes, based
on the multiuser GHZ channel. The main feature of the W-quantum
channels is a set of non-local operators, that allow receivers
recovering unknown state.
\end{abstract}


\maketitle


\section{Introduction}

Recently
a large
number of new entangled states has been proposed, particulary
the W-class, introduced by Cirac et al \cite{Cirac}, the states
called zero sum amplitude (ZSA) by Pati \cite{Pati}, a set of
four-particle entanglement considered by Verstraete et al
\cite{FourEn} and others. Although the features of these entangled
states are quite various each of them can be attractive for the
quantum information processing. In fact, it has been shown, that
ZSA-states are useful in the quantum web for preparing an
entanglement of an unknown qubit with two types of reference
states \cite{Pati}. Recently Lee et al \cite{Lee} have considered
W-states for secure communication.

It is well known, that the states of the GHZ-class and W-class are
rather different with respect to loss of qubits, so that the
W-channel looks more attractive because of its robustness. Now we
have a lot of entangled states and one of the main questions is
what informational tasks  could be achieved using a given
entangled state.
We consider this question for 
the case of the W-class states. The aim of the work is to
investigate teleportation of entangled states. It can be done by
different ways, particularly using the GHZ
(Greenberger-Horne-Zeilinger) channel, shared with a sender and
receivers \cite{AV}, \cite{Luca}, \cite{NGHZ}.

Our start point is the GHZ protocol which can be transformed to
the desired case by unitary operations, involved both the channel
and the state to be teleported. It is well known that the states
of the GHZ and W-class can't be converted from one to another by
local operations and classical communication (LOCC) \cite{Cirac}.
This fact results in the main feature of the W-class channel, when
receivers can recover an unknown state to be teleported by
non-local operations only. Because these operations don't involve
the unknown state the task is accomplished. Indeed, considering
how to teleport two-qubit state by the four-particle entanglement,
it was found by Lee et al \cite{LeeM}, that the recovering
operations can be non-local also. Any state from the GHG class is
not suitable as the quantum channel for teleportation. The reason
is that, it needs the states, that can be converted from the GHZ
triplet by two-particle unitary operation only, but it is
impossible for all members of the GHZ class. Also it is true for
the W-class and one finds a particular set of the states to be
useful.

The work is organized as follows. First we introduce two
transformations of the GHZ protocols that result in new channel
and new measurement. Then a set of states from the W-class are
considered as the quantum channels, next we establish new type of
entangled states that can be teleported using the GHZ channel and
discuss 
an optical implementation of the W-states.

\section{Two types of transformations}

We start from the GHZ channel represented by the three-qubit
entangled state of the form
\begin{equation}\label{100}
|GHZ\rangle=1/\sqrt{2}(|000\rangle+|111\rangle)_{ABC}.
\end{equation}
The teleportation protocol using (\ref{100})
as a channel allows transmitting a two-qubit entangled state like
EPR pair
\begin{equation}\label{102}
|A\rangle=(\alpha|01\rangle+\beta|10\rangle)_{12},
\end{equation}
where $|\alpha|^{2}+|\beta|^{2}=1$. The scheme includes five
particles in the total state, that is a product $|A\rangle_{12}
\otimes |GHZ\rangle_{ABC}$, where particles A, B and C of the
channel are shared with Alice and two receivers Bob and Claire,
spatially separated. If Alice decided sending the state
(\ref{102}) to Bob and Claire, she performs a measurement on three
particles 1,2 and A in the basis $\{\Phi_{x}\}$, that reads
\begin{equation}\label{103}
\{\Phi_{x}: \pi_{1}^{\pm}\otimes\Phi^{\pm}_{2A},
\pi^{\pm}\otimes\Psi^{\pm}_{2A}\},
\end{equation}
where $\Phi^{\pm}=(|00\rangle\pm|11\rangle)/\sqrt{2}$,
$\Psi^{\pm}=(|01\rangle\pm|10\rangle)/\sqrt{2}$ are the Bell
states,
$\pi_{1}^{\pm}=(|0\rangle\pm\exp(i\theta)|1\rangle)/\sqrt{2}$. The
measurement projects all particles into one of the eight states
$|\Phi_{x}\rangle_{12A}\otimes |BC_{x}\rangle_{BC}$ with equal
probabilities $Prob(\Phi_{x})$ to be independent from
$|A\rangle_{12}$, where $|BC_{x}\rangle_{BC}$ is a state in the
Bob and Claire hands. It results in successful teleportation
because there is a set of unitary operations $\{U_{x}\}$, we shall
name recovering operators, that recover the unknown state. Indeed
all operations can be chosen in the form of product
$U_{x}=B_{x}\otimes C_{x}$, where $B_{x}$ and $C_{x}$ acts to
particle $B$ and $C$ \cite{AV}. Or in more detail
\begin{eqnarray}\label{104}
|A\rangle_{12} \otimes |GHZ\rangle_{ABC}=&&
\\
\nonumber
=\sum_{x}|\Phi_{x}\rangle_{12A}\sqrt{Prob(\Phi_{x})}(B_{x}\otimes
C_{x})|A\rangle_{BC},&&
\end{eqnarray}
where $Prob(\Phi_{x})=1/8$ and $C_{x}, B_{x}$ are the Pauli operators.

There are two main resources, such as the set of recovering
operators and the measurement, that can be modified independently
by some transformations of the channel or the state to be
teleported.\\ Let $T$ be a two-particle unitary operator, that
converts the GHZ state into another one
\begin{equation}\label{106}
(1\otimes T)|GHZ\rangle_{ABC}=|\Omega\rangle.
\end{equation}
When this transformation is applied to (\ref{104}), it results in replacing of
the recovering operators
\begin{equation}\label{107}
B_{x}\otimes C_{x} \to T(B_{x}\otimes C_{x}),
\end{equation}
In the same time equations (\ref{106}) and (\ref{107}) tell that
to teleport entangled state (\ref{102}) a new channel $\Omega$ can
be used instead of the GHZ one and it needs new set of recovering
operators. Generally these two-particle operators are not
factorized or non-local that is a feature of the multiparticle
quantum channel. In spite of the operators can be non-local, the
task is accomplished, because they don't involve any qubits of the
state to be teleported. Here and later we use the simple notation,
that operators are local if $U|\varphi\rangle_{AB}=A\otimes
B|\varphi\rangle_{AB}$, where operator A acts on the particle A
and don't affect to B and so on.

Consider a two-particle unitary operator $R$, that converts the state to be
teleported
\begin{equation}\label{108}
R|A\rangle=|Z\rangle.
\end{equation}
This operator $R$ transforms (\ref{104}) in such a way that
instead of the basis $\Phi_{x}$, one finds
\begin{equation}\label{109} |\Phi_{x}\rangle\to
R|\Phi_{x}\rangle.
\end{equation}
It follows from (\ref{108}) (\ref{109}), that  new state $Z$ can
be teleported by the GHZ channel, when new measurement is
introduced.

Two considered transformations result in a simple observation. \emph{If the GHZ
channel allows teleportation of a state $A$, then any state to be unitary
equivalent to $A$, can be transmitted successfully, also the unknown state $A$
can be teleported  using any channel, obtained from the GHZ one by unitary
operation, that involves all particles of the channel except one}.

Note, the GHZ channel can be used for dense coding, when a three
bit message is transmitted by manipulating two bit of the channel
only \cite{NGHZ}, \cite{JCer}. It is possible because of there is
a set of the two-particle unitary operators $\{U'_{x}\}$, that
generate a complete basis of entangled states $\{\Phi_{x}'\}$ from
one of them, say from the GHZ one. In other words 3 bits of
classical information are encoded by the set $\Phi_{x}$ and a
three-particle measurement allows to extract the message. All
operators can be chosen in the factorized form
$U_{x}'=B'_{x}\otimes C'_{x}$, and
\begin{equation}\label{110}
|\Phi'_{x}\rangle=(1\otimes B'_{x}\otimes
C'_{x})|GHZ\rangle_{ABC}.
\end{equation}
This scheme can be modified in  
the same way, when a new channel or new measurement is introduced.

What kind of states from the GHZ class can be chosen as a quantum channel for
teleportation or dense coding schemes? One of the subsets of the three-particle
states can be obtained using equations (\ref{106}) or its particular case
(\ref{110}). By this way one finds a collection
\begin{equation}\label{11010}
|\Phi'_{x}\rangle=1/\sqrt{2}(|0\rangle|x\rangle\pm|1\rangle|\overline{x}\rangle).
\end{equation}
where $x=00,01,10,11$, that is a complete basis. In the same time any
normalized state of the form $a|000\rangle+b|111\rangle$, where $|a|\neq |b|$
is insufficient as the quantum channel. For teleportation it results from the
fact, that measurement, given by (\ref{103}), has outcomes, which probabilities
depend from the unknown state $(\ref{102})$ as
$Prob=1/8[1\pm(|a|^{2}-|b|^{2})(|\alpha|^{2}-|\beta|^{2})]$. For dense coding
it follows from the fact, that in accordance with (\ref{110}) one can generate
a set of states $a|0\rangle|x\rangle\pm b|1\rangle|\overline{x}\rangle$, but
the set is not a complete basis.

Indeed the state of the GHZ class
\begin{equation}\label{11011}
|f_{GHZ}\rangle=1/\sqrt{3}(\sqrt{2}|000\rangle+|111\rangle)
\end{equation}
can't be used as a quantum channel. It is in agreement with
equation (\ref{106}), that tells, that $T$ must be a unitary
operator. The reason is that any two states from the GHZ class
can't be converted from one to another by unitary transformation,
if a two-particle operator is permitted only. It is true for the
$|GHZ\rangle$ and $|f_{GHZ}\rangle$ state. In fact, consider the
transformation
\begin{equation}\label{11012}
  (1\otimes T)(a|000\rangle+b|111\rangle)
  =a'|000\rangle+b'|111\rangle.
\end{equation}
Because of $T$ is the unitary operator it results in conditions
$|a|=|a'|$ and $|b|=|b'|$, then there is no unitary
transformation, that converts $|GHZ\rangle$ to $|f_{GHZ}\rangle$.

In other words the states of the GHZ class have different features
under two-particle transformation when only a subset of the class
can be used as the quantum channel for teleportation and dense
coding. Indeed this observation is true for the W-class.

\section{The channel of the W-class}

Consider a particular case of transformation $T$, given by
(\ref{106}), that converts the GHZ state into a state of the
W-class. It follows from the result of Cirac et al \cite{Cirac},
that the operator $T$ is non-local. Now we are interested in two
states from the W-class only, say of the form
\begin{eqnarray}\label{200}
|W\rangle=1/\sqrt{3}(|100\rangle+|010\rangle+|001\rangle)&&\\
|\widetilde{W}\rangle=1/\sqrt{2}(|100\rangle+|0\Psi^{+}\rangle).&&
\end{eqnarray}

Introduce the unitary non-local operator $V$
\begin{eqnarray}\label{203}
V=|\Psi^{+}\rangle\langle 00|+ |11\rangle\langle 01|
+|\Psi^{-}\rangle\langle 10|+|00\rangle\langle 11|.&&
\end{eqnarray}
It can be represented by network, including the conditional gates
$V=C_{12}(C-H)_{21}\sigma_{x2}C_{21}$, where $C_{ct}$ is $C-NOT$
operation, $c$ is controlled bit, $t$ is target one, $(C-H)_{21}$
is the controlled Hadamard gate, that transforms
$|b0\rangle\to|b0\rangle$, $|b1\rangle\to (H\otimes 1)|b1\rangle$,
$b=0,1$. The operator $V$  converts $00\to\Psi^{+}$, $01\to 11$,
$10\to\Psi^{-}$, $11\to 00$ and generates a complete set of states
$\Psi^{\pm}, |00\rangle, |11\rangle$ that was used by Basharov to
consider the problem of atomic relaxation under the entangled
thermostat\cite{Askhat}.

The introduced unitary operator $V$ converts the states from the
GHZ class into the W-class
\begin{eqnarray}\label{2030}
(1\otimes V)|GHZ\rangle_{ABC}=|\widetilde{W}\rangle&&\\
(1\otimes V)|f_{GHZ}\rangle_{ABC}=|W\rangle.&&
\end{eqnarray}
Because of two states $|GHZ\rangle$ and $|f_{GHZ}\rangle$ are
inequivalent under two-particle transformation, the obtained
states of the W-class have also their properties to be different.
Clear, that in contrast $|W\rangle$, the $|\widetilde{W}\rangle$
state is sufficient as a quantum channel for teleportation of
entangled states and dense coding. As well a collection $(1\otimes
V)|\Phi'_{x}\rangle$, including $|\widetilde{W}\rangle$, where
$\Phi'_{x}$ is given by (\ref{11010}), represents  a set of the
quantum channels of the W-class. For teleportation it results in
new feature of recovering operators which become non-local ones.

The original state of the W-class in the standard unique form reads
\cite{Cirac}
\begin{equation}\label{204}
\varphi_{W}=\sqrt{a}|001\rangle+\sqrt{b}|010\rangle+\sqrt{c}|100\rangle+\sqrt{d}|000\rangle,
\end{equation}
where $a,b,c >0$ and $d=1-a-b-c\geq 0$. Considering $\varphi_{W}$
as a quantum channel, there is a question whether teleportation of
the two qubits in the general state is successful. The answer is
not because of the following reasons. Let a total state be a
product $|A'\rangle_{12}\otimes |\varphi_{W}\rangle_{ABC}$, where
$|A'\rangle=(\gamma|00\rangle+
\alpha|01\rangle+\beta|10\rangle+\delta|11\rangle)_{12}$ is a
general two-qubit state. When the measurement in the basis, given
by (\ref{103}), performs, all probabilities of outcomes depend on
the unknown state $A'$
\begin{eqnarray}\label{205}
Prob(\pi^{\pm}\Phi^{\pm})
 =[|\gamma\pm\exp(i\theta)\beta|^{2}(1-a)&&\\
\nonumber
 +a|\alpha\pm\exp(i\theta)\delta|^{2}]/4&&\\
 \nonumber
Prob(\pi^{\pm}\Psi^{\pm})
 =[|\gamma\pm\exp(i\theta)\beta|^{2}a&&\\
\nonumber
 +(1-a)|\alpha\pm\exp(i\theta)\delta|^{2}]/4.&&
\end{eqnarray}
However the task can be accomplished for the considered particular
cases. Indeed, it is clear without any calculations from the
following reasons. From the physical point of view any
two-particle entangled state looks more as one piece of reality.
To teleport a qubit or one piece, it needs  an EPR channel of two entangled
particles, where one of them is involved in 
measurement and another particle is a blank. For teleportation of two entangled
qubits it needs a channel, including two particles as blanks and one particle
for measurement, then one finds a three-particle channel at last. It is not
true, when two qubits are not entangled.

\section{The states teleported by the GHZ-channel}

In accordance with (\ref{108}) and (\ref{109}) a unitary
transformation $R$ allows to investigate, what kind of states
could be teleported using the GHZ-channel.

Let $R=V$, for example. Then
\begin{equation}\label{301}
V|A\rangle=\alpha|11\rangle+\beta|\Psi^{-}\rangle.
\end{equation}
To teleport this entangled state, it needs a new three-particle measurement in
the basis, obtained from (\ref{103}), when $ |\pi^{+}\Phi^{\pm}\rangle
\leftrightarrow 1/\sqrt{2}(|010\rangle\pm|\Phi^{\pm}\rangle|1\rangle) $ and so
on. The found set is represented by the states of the W-class.

Consider the GHZ channel of four qubits
\begin{equation}\label{304}
|GHZ\rangle=1/\sqrt{2}(|0000\rangle+|1111\rangle)_{ABCD}.
\end{equation}
Let particles $A$, $B$, $C$ and $D$ share a sender Alice and
three receivers $B$, $C$, $D$. It has been found, that a
three-particle state of the form
$|A\rangle_{123}=\alpha|000\rangle+\beta|111\rangle$ can be
teleported, when Alice performs a joint measurement, described by
the set $\{\Phi_{x}:
\pi_{1}^{\pm}\otimes\pi_{2}^{\pm}\otimes\Phi^{\pm}_{3A};~
\pi_{1}^{\pm}\otimes\pi_{2}^{\pm}\otimes\Psi^{\pm}_{3A}\}$
\cite{AV}. A simple observation shows, that the channel is useful
for transmitting a state of the W-class. In fact, it follows from
(\ref{203}), that
\begin{equation}\label{305}
(1\otimes V)(\alpha
|000\rangle+\beta|111\rangle)=\alpha|100\rangle+\beta|0\Psi^{+}\rangle.
\end{equation}
Then the useful teleportation of entangled state
$\alpha|100\rangle+\beta|0\Psi^{+}\rangle$ is obtained.

Consider one of the way how to prepare the states of the W-class
for particular case of optical implementation. Let a collection of
two-level identical atoms interacts with a set of modes, which
frequencies are close to frequency of an atomic transition for
simplicity. The density matrix of atoms and field $F$ obeys
equation of the form
\begin{equation}
\partial F/\partial t=-i\hbar^{-1}[V, F],
\end{equation}
where
$V=SB^{\dagger}+S^{\dagger}B$
is Hamiltonian, $S,S^{\dagger}$ are atomic operators,
$B=g_{1}b_{1}+g_{2}b_{2}+g_{3}b_{3}$;   $b_{k}, b_{k}^{\dagger},
k=1,2,3$ are the photon operators of modes. Let atoms have
coherence, in the sense, that their polarization $\langle
S\rangle\neq 0$. It can be done, if atoms are illuminated by a
strong coherent wave of amplitude $\alpha$, then $\langle S\rangle
=q\alpha$. One can describe behavior of the field by a master
equation for density matrix of modes $\rho=Sp_{atoms}F$. In the
first non-vanishing order the master equation for $\rho$ takes the
form
\begin{equation}
\partial \rho/\partial t=-i\hbar^{-1}[V', \rho],
\end{equation}
where an effective Hamiltonian $V'=q(\alpha^{\ast}B+\alpha
B^{\dagger})$ is obtained. In fact, it describes a parametric
scattering of the strong wave into three week modes and can be
rewritten as
\begin{equation}
\partial |\Psi\rangle/\partial t=-i\hbar^{-1}V'|\Psi\rangle.
\end{equation}
 Let the state of the week modes be vacuum, then in the first
order over interaction one finds
\begin{equation}
|\Psi(t)\rangle=|000\rangle-i\hbar^{-1}\alpha t|w\rangle,
\end{equation}
where
$|w\rangle =g_{1}|100\rangle+g_{2}|010\rangle+g_{3}|001\rangle$,
is a state of the W-class.

\section{Conclusion}

Being attractive because of its robustness with respect to losses
of particle the W-states can be used for the teleportation and
dense coding tasks. In the same time the states from the W-class
are different under the unitary two-particle transformations and
only some of them can be converted from one to another by this
way. It results in a subset of the states to be suitable only. We
find the desired collection of states using unitary transformation
of the GHZ protocols. The W-channel has new feature, that is
non-local operators recovering unknown state teleported.

We are grateful to Alexander Yu. Vlasov for discussion. This work
was supported in part by the Delzell Foundation Inc. and INTAS
grant 00-479.


\begin{thebibliography}{99}
\bibitem{Cirac} W. Dur, G. Vidal, J.I. Cirac. Phys. Rev. A, 62,
062314, (2000). Three qubits can be entangled in two inequvalent ways. E-print,
LANL, quant/ph 0005115.
\bibitem{Pati} A.K. Pati. Quantum Cobweb: Multiparty Entangling of an unknown Qubit.
E-print, LANL, quant/ph 0101049.
\bibitem{FourEn}  F. Verstraete, J. Dehaene, B. De Moore, H. Verscheld.
Four qubits can be entangled in nine different ways. E-print,
LANL, quant/ph 0109033.
\bibitem{Lee} J. Joo, Y-J Park, J. Lee, J. Jang. Quantum Secure
Communication with W States. E-print, LANL, quant/ph 0204003.
\bibitem{AV} V.N. Gorbachev, A.I. Trubilko. JETP, 118, 1036, (2000). Quantum
teleportation of EPR pair by three particle entanglement. E-
print, LANL, quant/ph 9906110.\\
V.N. Gorbachev, A.I. Trubilko, A.I. Zhiliba. J. Opt. B, 3, S25,
(2001).
\bibitem{Luca} L. Marinatto, T. Weber.  Found. Phys. Lett., 13,
119, (2000). Which kind of two-particle state can be teleported
throught a three-particle quantum channel. E-print, LANL,
quant/ph 0004054.
\bibitem{NGHZ}
V.N. Gorbachev, A.I. Zhiliba, A.I. Trubilko, E.S. Yakovleva.
Teleportation of entangled states and dense coding using a
multiparticle quantum channel. E-print, LANL, quant/ph 0011124.
\bibitem{LeeM} J. Lee, H. Min, S.D. Oh. Multiparticle entanglement
for an teleportation. E-print, LANL quant/ph 0201069.
\bibitem{JCer} J.L. Cereceda. Quantum dense coding using three
qubits. E-print, LANL quant/ph 0105096.
\bibitem{Askhat} A.M. Basharov. Pis'ma v ZhETF  75, 151-154 (2002).

\end{thebibliography}
\end{document}